\documentstyle{ichep}
\begin{document}

\title{
\vskip -2.8cm
\hfill {\normalsize TTP94--19}\\
\vskip -0.2cm
\hfill {\normalsize September 1994}\\
\vskip 1.0cm
QCD Corrections to Inclusive Distributions of Leptons\\
in Decays of Polarised Heavy Quarks*}
\author{
\vskip-0.5cm
Marek Je\.zabek}

\affil{Institute of Nuclear Physics,
        ul.Kawiory 26a, PL-30055 Cracow, Poland\\
and\\
Institut f. Theoretische Teilchenphysik, Univ. Karlsruhe
D-76128 Karlsruhe, Germany}
\abstract{
Compact analytic expressions have been obtained for the first order
perturbative QCD corrections to the inclusive spectra of the leptons
in the semileptonic decays of polarised heavy quarks.
\cabs
Charmed and beautiful $\Lambda$ baryons from $Z^0$ decays can be viewed
as sources of highly polarised charm and bottom quarks. Charged leptons
and neutrinos from $\Lambda_b$ and $\Lambda_c$ decays can be used in
the polarisation studies for the corresponding heavy quarks. Thus our
results are applicable for the b quark polarisation measurements at LEP.
\cabs
Short lifetime  enables polarisation studies for the top quark.
The angular-energy spectra
of the charged leptons are particularly useful in this respect whereas
the distributions of the neutrinos are sensitive to deviations from
the V-A structure of the charged weak current in the decay.
}

\twocolumn[\maketitle]

\fnm{7}{Presented at XXVII International Conference on High Energy
Physics, 20-27 July 1994, Glasgow, Scotland; to appear in the
proceedings.}

\section{Introduction}
Inclusive semileptonic decays
of polarised charm and bottom quarks play important
role in present day particle physics. With
increasing statistics at LEP and good prospects for $B$-factories
quantitative description of
these processes may offer the most
interesting tests of the standard quantum theory
of particles. In fact the first measurement of $b$ quark
polarisation at LEP has been presented by the ALEPH collaboration
at this conference \cite{Roudeau}.
At the high energy frontier semileptonic decays
of the top quark will be instrumental in establishing its
properties \cite{Kuehn1,Kuehn2,Kuehn3,teupitz}.

In this article I present the results of calculations
of the first order perturbative
QCD corrections to semileptonic
decays of polarised heavy quarks. Some of these results
have been published in \cite{CJK,CJKK}.
In \cite{CJ} compact analytic formulae have been obtained
for the distributions of the charged lepton and the neutrino.
These formulae  agree with those given in \cite{CJK}
for the joint angular and energy distribution
of the charged lepton in top quark decays and are much simpler.

\section{The formula and cross checks}
The QCD corrected triple differential distribution
of the charged lepton
for the semileptonic decay  of the polarised quark
with the weak isospin $I_3=\pm 1/2$
can be written in the following way \cite{CJ}:
\begin{eqnarray}
{{\rm d}\Gamma^{\pm} \over {\rm d}x\,{\rm d}y\,{\rm d}\cos\theta  }
&& \sim\qquad \left[\,
{\rm F}^\pm_0(x,y) + S\cos\theta\,{\rm J}^\pm_0(x,y)\,
\right]
\nonumber\\
&& -\; {2\alpha_s\over3\pi}\;
  \left[\,
  {\rm F}^\pm_1(x,y) + S\cos\theta\,{\rm J}^\pm_1(x,y)\,
  \right]
\nonumber\\
\label{eq:1}
\end{eqnarray}
In the rest frame of the decaying heavy quark
$\theta$ denotes the angle between the
polarisation vector $\vec s$ of the heavy quark and the
direction of the charged lepton,
$S=|\,\vec s\,|$,
$x= 2Q\ell/Q^2$ and $y= 2\ell\nu/Q^2$ where
$Q$, $\ell$ and $\nu$ denote the four-momenta of the decaying
quark, charged lepton and neutrino. Eq.(\ref{eq:1}) describes also
the triple differential
distribution of the neutrino for  $I_3=\mp 1/2$. In this case,
however, $x= 2Q\nu/Q^2$  and
$\theta$ denotes the angle between
$\vec s$ and the three-momentum of the neutrino.
The functions ${\rm F}^\pm_0(x,y)$ and ${\rm J}^\pm_0(x,y)$
corresponding to Born approximation read:
\begin{eqnarray}
{\rm F}^+_0(x,y) &=& x (x_m-x)
\label{eq:5}\\
{\rm J}^+_0(x,y) &=& {\rm F}^+_0(x,y)
\label{eq:6}\\
{\rm F}^-_0(x,y) &=& (x-y) (x_m-x+y)
\label{eq:7}\\
{\rm J}^-_0(x,y) &=& (x-y) (x_m-x+y-2y/x)
\label{eq:8}
\end{eqnarray}
where $x_m=1-\epsilon^2$, $\epsilon^2= q^2/Q^2$ and $q$
denotes the four-momentum of the quark originating from
the decay.
The functions ${\rm F}^\pm_1(x,y)$ and
${\rm J}^+_1(x,y)$ correspond to
the first order QCD corrections and
are given in \cite{CJ}.

Non-trivial cross checks are fulfilled by the polarisation
independent parts of the distributions (\ref{eq:1}):
\begin{itemize}
\item
the distributions
${\rm d}\Gamma^{\pm} / {\rm d}x\,{\rm d}y$
agree with the results
for unpolarised decays which were
obtained in \cite{JK}. The present formulae are simpler.
\item
in the four-fermion (Fermi) limit integration over $y$
can be performed numerically. The resulting distributions
${\rm d}\Gamma^{\pm} / {\rm d}x$
also agree with those of \cite{JK}.
Recently the results of \cite{JK}
have been confirmed \cite{CCM1}. Thus an old conflict
with other calculations \cite{CCM0} is solved and
the agreement with \cite{JK} can be considered as a
non-trivial cross check. Moreover, the
analytic result of \cite{JK} for
${\rm d}\Gamma^{+} / {\rm d}x$
and $\epsilon=0$
has been also confirmed \cite{GWF}.
\item
$${{\rm d}\Gamma^{+} / {\rm d}y} =
{{\rm d}\Gamma^{-} / {\rm d}y}$$
and the analytic formula for this distribution
exists \cite{JK0} which at the same time describes the lifetime
of the top quark as a function of its mass.
This formula has been confirmed by a few groups, c.f. \cite{JK93}
and references therein.
\item
in the four-fermion limit the result for the total rate $\Gamma$
derived from
eq.(\ref{eq:1})
agrees with the results of \cite{CM} and
the analytical formula of \cite{nir}.
\end{itemize}

\section{Applications}
\subsection{Polarised bottom and charm quarks}

Polarisation studies for heavy flavors at LEP are a new
interesting field of potentially fundamental significance,
see \cite{Mele,Roudeau} for recent reviews.
According to the Standard Model
$Z^0\to b\bar b$ and
$Z^0\to c\bar c$ decays
can be viewed as sources of highly polarised heavy quarks.
The degree of longitudinal polarisation is fairly large,
amounting to $\langle P_b\rangle = -0.94$ for $b$
and $\langle P_c\rangle = -0.68$ for $c$ quarks~\cite{Kuehn1}.
The polarisations depend weakly on the production angle.
QCD corrections to Born result are about 3\% \cite{KPT}.
The real drawback is that due to hadronisation the net longitudinal
polarisation of the decaying  $b$ and $c$ quarks is drastically
decreased. In particular these $b$ quarks become depolarised
which are bound in $B$ mesons
both produced directly and from $B^*\to B\gamma$ transitions.
The signal is therefore significantly reduced.
Only those $b$'s (a few percent) which fragment directly
into $\Lambda_b$  baryons retain information on the original
polarisation \cite{Bjo}.
Polarisation transfer from a heavy
quark $Q$ to the corresponding $\Lambda_Q$ baryon is 100\%
\cite{CKPS} at least in the limit $m_Q\rightarrow\infty$.
Thus, a large net polarisation is expected for heavy quarks
in samples enriched with these heavy baryons.

It has been proposed long ago \cite{zerwas}
that distributions of charged leptons from semileptonic
decays of beautiful hadrons can be used in polarisation
studies for $b$ quarks.
Some advantages of neutrino distributions have been
also pointed out \cite{CJKK,BR,dittmar}.
Recently there has been considerable progress in the theory of
the inclusive semileptonic decays of heavy flavor hadrons.
It has been shown that in the leading order of an expansion
in inverse powers of heavy quark mass $1/m_Q$ the
spectra for hadrons coincide with those for the decays of
free heavy quarks \cite{CGG} and there are no $\Lambda_{QCD}/m_Q$
corrections to this result away from the energy
endpoint.
$\Lambda^2_{QCD}/m_Q^2$ corrections have been
calculated in \cite{Bigi,wise} for $B$ mesons and
in \cite{wise} for polarised $\Lambda_b$ baryons.
For some decays the results are similar to those of the well-known
$ACCMM$ model \cite{altar}.
The corrections to charm decays are larger than for bottom
and convergence of
$1/m_Q$  expansion is poorer \cite{Shifman}.
Perturbative first order QCD corrections contribute 10-20\%
to the semileptonic decays and for bottom are much larger
than the nonperturbative ones.

\subsection{Polarised top quarks}
The analysis of polarised top quarks and their decays
has recently attracted considerable attention,
see \cite{Kuehn3,teupitz} and references
cited therein.
The reason is that this analysis
will result in determination of the top
quark coupling to the $W$ and $Z$ bosons either
confirming the predictions
of the Standard Model
or providing clues for physics beyond.
The latter possibility is particularly intriguing for the top quark
because $m_t$
plays an exceptional role in the fermion mass spectrum.

A number of mechanisms have been suggested that will
lead to polarised top quarks.
Studies at a linear electron-positron collider are particularly
clean for precision tests. However, also
$\gamma\gamma$ collisions with circular polarised photons
and subsequent spin analysis of top quarks might reveal new
information.
Related studies may be performed in hadronic collisions
which in this case are mainly based on the correlation
between $t$ and $\bar t$ decay products.
However, single top production through $Wb$ fusion at LHC
may also be a useful source of polarised top quarks.
Electron-positron collisions are
the most efficient and flexible
reactions producing
polarised top quarks. A small
component of polarisation transverse to the production plane
is induced by final state interactions.
The longitudinal polarisation $P_L$
is large.  $P_L$ varies strongly with the production angle.
Averaging over the production angle
leads therefore to a significant reduction of $P_L$ with typical
values of $\langle P_L\rangle$ around -0.2 \cite{KPT}.\\
All these reactions lead to sizable polarisation and can be used
to obtain information on the production mechanism.
However, two drawbacks are evident:
production and decay are mixed in an intricate manner, and
furthermore the degree of polarisation is relatively small
and depends on the production angle. Top quark production
with longitudinally polarised electron beams and close to
threshold provides one important exception: the restricted
phase space leads to an amplitude which is dominantly S-wave
such that the electron (and positron) spin is directly
transferred to the top quark. Close to threshold and with
longitudinally polarised electrons
one can study decays of polarised top quarks
under particularly convenient conditions:
large event rates, well identified rest frame of the top quark,
and large degree of polarisation.  Moreover, short lifetime
of top quark practically eliminates nonperturbative corrections
due to hadronisation.

In the rest frame of the decaying $t$ quark distributions
of the decay products are sensitive to its polarisation.
Eq.(\ref{eq:6}) implies that in Born approximation
the double differential angular-energy distribution
of the charged lepton is the product of the energy
distribution and the angular distribution. The latter distribution
is of the following form
\begin{equation}
{ {\rm d}N\over{\rm d}\cos\theta} =
{1\over 2}\, \left[\, 1\,+
\,S\cos\theta \right]
\label{eq:elec1}
\end{equation}
QCD corrections essentially do not spoil factorisation
of the charge lepton distribution \cite{CJK}.
It is noteworthy that for $S$=1 the angular
dependence in (\ref{eq:elec1}) is maximal because
any larger coeffecient multiplying $\cos\theta$ would be
in conflict with positivity of the decay rate.
Thus the polarisation analysing power of
the charged lepton energy-angular distribution
is maximal \footnote{This is reversed
for $b$ decays and the polarisation analysing power is maximal
for the neutrino distributions
because the formulae for the neutrino distributions
in down-type quark decay describe the charged lepton distributions
for an up-type quark.}.

It follows from eqs.
(\ref{eq:7}) and (\ref{eq:8}) that already in Born approximation
there is no factorisation
for the neutrino energy-angular distribution.
Neutrino distributions are therefore less sensitive to the
polarisation of the decaying top quark than charge lepton
distributions
On the other hand it has been shown \cite{JK94} that
the angular-energy distribution of neutrinos
from the polarised top quark decay will allow for a particularly
sensitive test of the V-A structure of the weak charged current.
The effect of QCD correction can mimic a small admixture
of V+A interaction. Therefore, inclusion of the radiative
QCD correction to the decay distributions  is necessary for
a quantitative study.

\vskip0.5cm\noindent
{\bf Acknowledgements}\\
I thank  Andrzej Czarnecki, Hans K\"uhn
and J\"urgen K\"orner for collaborations
on research reported in this article.
I would like to gratefully acknowledge helpful
correspondence with Professors N. Cabibbo,
G. Corbo and L. Maiani.
\par\noindent
This
work is partly supported by KBN under contract
2P30225206 and by DFG under contract 436POL173193S.


\Bibliography{9}
\bibitem{Roudeau} P.\ Roudeaud, {\it Heavy Quark Physics}, in
these proceedings.
\bibitem{Kuehn1}
J.H. K\"uhn and P.M. Zerwas, in {\it Heavy Flavours}, eds. A.J. Buras
and M. Lindner, (World Scientific, Singapore, 1992), p.434.
\bibitem{Kuehn2}
J.H. K\"uhn et al., DESY Orange Report 92-123A (1992), vol.I,p.255.
\bibitem{Kuehn3}
J.H. K\"uhn, "Top Quark at a Linear Collider", in {\it Physics
and Experiments with Linear $e^+e^-$ Colliders}, eds. F.A. Harris
et al., (World Scientific, Singapore, 1993), p.72.
\bibitem{teupitz}
M. Je\.zabek, {\it Top Quark Physics}, in proceedings
of Zeuthen workshop {\it Physics at LEP 200 and Beyond}, to appear
in \np{B}{94}{Suppl.};\ Karlsruhe\ preprint TTP94-09.
\bibitem{CJK}
A.~Czarnecki, M.~Je{\.z}abek and J.H. K{\"u}hn,
\np{B351}{91}{70}.
\bibitem{CJKK}
A. Czarnecki, M. Je\.zabek, J.G. K\"orner and J.H. K{\"u}hn,
\prl{73}{94}{384}.
\bibitem{CJ}
A. Czarnecki and M. Je\.zabek,
preprint TTP 93-40, Karlsruhe, 1994, hep-ph/9402326,
\np{B}{94}{in print}.
\bibitem{JK}
M.~Je{\.z}abek and J.H. K{\"u}hn,
\np{B320}{89}{20}.
\bibitem{CCM1}
N. Cabibbo, G. Corbo and L. Maiani, private communication.
\bibitem{CCM0}
N. Cabibbo, G. Corbo and L. Maiani,
\np{B155}{79}{93};\
G. Corbo, \np{B212}{83}{99}.
\bibitem{GWF}
C. Greub, D. Wyler and W. Fetscher,
\pl{B324}{94}{109}.
\bibitem{JK0}
M.~Je{\.z}abek and J.H. K{\"u}hn,
\np{B314}{89}{1}.
\bibitem{JK93}
M. Je\.zabek and J.H. K\"uhn, \prev{D48}{93}{R1910}.
\bibitem{CM}
N. Cabibbo and L. Maiani,
\pl{B79}{78}{109}.
\bibitem{nir} Y. Nir,
\pl{B221}{89}{184}.
\bibitem{Mele}
B. Mele, preprint n.1009, Rome, 1994.
\bibitem{KPT}
J.G. K\"orner, A. Pilaftsis and M.M. Tung, preprint MZ-TH/93-3,
\zp{C}{94}{in print}.
\bibitem{Bjo}
J.D. Bjorken, \prev{D40}{89}{1513}.
\bibitem{CKPS}
F.E. Close, J.G. K\"orner, R.J.N. Phillips and
D.J. Summers, \jpg{18}{92}{1716}.
\bibitem{zerwas}
G. K\"opp, L.M. Sehgal and P.M. Zerwas,
\np{B123}{1977}{77};\\
B. Mele and G. Altarelli, \pl{B299}{93}{345}.
\bibitem{BR}
G. Bonvicini and L. Randall,
\prl{73}{94}{392}.
\bibitem{dittmar}
M. Dittmar and Z. W{\c a}s,
\pl{B332}{94}{168};\\
M. Dittmar, in these proceedings.
\bibitem{CGG}
J. Chay, H. Georgi and B. Grinstein,
\pl{B247}{90}{399}.
\bibitem{Bigi}
I.~Bigi, M.~Shifman, N.~Uraltsev and A.~Vainshtein,
\prl{71}{93}{496}.
\bibitem{wise}
A.V. Manohar and M.B. Wise, \prev{D49}{94}{1310}.
\bibitem{altar}
G. Altarelli et al, \np{B208}{82}{365}.
\bibitem{Shifman}
M. Shifman, in these proceedings.
\bibitem{JK94}
M. Je{\.z}abek and J.H. K{\"u}hn,
\pl{B329}{94}{317}.
%
\end{thebibliography}
\end{document}